# Tailoring absorption in metal gratings with resonant ultra-thin bridges

M. A. Vincenti[1,*], D. de Ceglia[1], M. Grande[2], A. D'Orazio[2], M. Scalora[3]

[1] National Research Council - AMRDEC, Charles M. Bowden Research Laboratory, Redstone Arsenal - AL, 35898 – USA

[2] Dipartimento di Ingegneria Elettrica e dell'Informazione (DIEI), Politecnico di Bari, Via Re David 200, 70126 Bari – Italy

[3] Charles M. Bowden Research Laboratory, AMRDEC, US Army RDECOM, Redstone Arsenal - AL, 35898, USA

[*] e-mail: maria.vincenti@us.army.mil

**Abstract**

We present a theoretical analysis of the effects of short range surface plasmon polariton excitation on sub-wavelength bridges in metal gratings. We show that localized resonances in thin metal bridges placed within the slit of a free-standing silver grating dramatically modify transmission spectra and boost absorption regardless of the periodicity of the grating. Additionally, the interference of multiple localized resonances makes it possible to tailor the absorption properties of ultrathin gratings, regardless of the apertures' geometrical size. This tunable, narrow-band, enhanced-absorption mechanism triggered by resonant, short range surface plasmon polaritons may also enhance nonlinear optical processes like harmonic generation, in view of the large third-order susceptibility of metals.

## 1. Introduction

The transmission properties of metal gratings [1-3], including the verification and interpretation of the enhanced transmission effect [4,5], have inspired a large number of studies [6-8]. Using diverse geometrical patterns, sizes and materials, researchers have extended the theory of enhanced transmission across the optical spectrum and beyond [9-15], suggesting that gratings with sub-wavelength features may be crucial for several applications such as sensing [16,17], switching [18] and filtering [19-21]. Surface plasmon (SP) excitation and the consequent, dramatic increase of the electromagnetic field density in sub-wavelength regions also suggest the use of periodically structured metals for enhanced Raman scattering applications [22-25] and harmonic generation from the visible down to UV and soft X-rays regimes [17,26-31].

In parallel with transmission property studies, several groups have also focused their attention on extraordinary optical absorption because it may easily exceed the performance of enhanced transmission properties, and because of its potential applications to ultra-small detectors [32,33], efficient thermal emitters [34], and energy harvesting devices [35]. In this regard, metal gratings composed of sub-wavelength grooves have been widely investigated. The excitation of SPs in these structures is allowed by the modification of the metallic surface at the nanometer scale. However, shallow and deep trenches may behave in very different ways. For example, while the former supports the excitation of SPPs in the quasi-static regime [36], allowing for nearly perfect absorption, the latter support transverse electric magnetic (TEM) modes [37], whose resonant conditions and increased absorption are strictly related to the geometry of the groove. Properly designed, grooved metal films can also bring light to a virtual stand-still [38]. This trapping mechanism occurs for frequencies in the vicinity of the plasmonic band edge, where the group velocity of the surface modes is significantly reduced[38], and may be extended over a broad frequency range by carefully engineering the 1D-periodic profile of the nanostructure.

Since most groove-related efforts have been carried out by considering corrugations on either relatively thick or semi-infinite metal layers [39-44], the effects of the excitation of SPs are usually inferred by observing reflection spectra, as little or no attention is paid to the effects of corrugations on transmission properties. In fact, when corrugated thin [45] and thick [46] layers are considered one can appreciate the presence of SP modes by also observing changes that occur in the respective transmission spectra. In a zero-order grating with sinusoidal corrugation, maxima/minima in transmission/reflection may be related to the tunneling of localized SPs trapped on the grating [46]. This tunneling effect is somewhat insensitive to incident angle and thus adequate for optical filtering purposes.

In this paper we investigate a different type of resonant mechanism that may also be useful for optical filtering and other applications that require engineered absorption profile. Differently from the structure shown in Ref. [46], where the tunneling regime increases transmission and decreases reflection, here we exploit the excitation of localized resonances to abate both transmission and reflection in favor of increased absorption. This subtle difference makes these structures suitable for sensor and/or eye protection applications. We show that ultra-thin metal bridges placed in the midst of thick or thin gratings may support resonant, short range SP modes that dramatically alter transmission and absorption spectra. We first investigate the dependence of resonant modes supported by the metal bridges as a function of the geometrical parameters, i.e. aperture size, aperture depth and bridge thickness, for different grating thicknesses, in order to understand the mechanism to achieve efficient narrow-band filtering by enhancing absorption. The results reveal that the excitation of these modes modulates absorption and transmission spectra in the visible and near infrared (NIR) ranges. However, the effects of these localized resonances may be dramatically different depending on grating thickness. While for thick grating resonances localized in the ultra-thin bridge only enhance absorption values, we show that for thin gratings the interplay between resonant modes with different parity leads to anti-crossing and absorption features that are virtually independent of aperture size and position of the metal bridge inside the slit. The increased absorption regime induced by strong field localization in the ultra-thin metal bridges may be exploited to trigger nonlinear processes that arise from the large, third order bulk nonlinearity of metals.

## 2. Surface plasmons on insulator/metal/insulator systems

Surface plasmon propagation on metal layers surrounded by dielectric materials has been extensively studied [47,48], together with dispersion properties and field distributions [49-51]. For example, it has been shown that there is substantial difference between surface plasmons propagating on thick and thin metal surfaces. A thick metal layer (i.e., thickness much greater than the plasmon skin depth) can sustain decoupled surface waves at each metal/dielectric interface [47]. In contrast, for thin metal layers one can no longer neglect the interaction between the waves propagating on the two surfaces [47,52]. As a consequence, the SP dispersion curve associated with a metal/dielectric interface splits into two branches describing high- ($\omega^+$) and low-frequency ($\omega^-$) modes, whose dispersion relations are [47]:

$$\omega^+ : \quad \varepsilon_1 k_{z2} + \varepsilon_2 k_{z1} \tanh\left(\frac{-ik_{z1}d}{2}\right) = 0 \quad , \quad (1)$$

$$\omega^- : \quad \varepsilon_1 k_{z2} + \varepsilon_2 k_{z1} \coth\left(\frac{-ik_{z1}d}{2}\right) = 0 \quad , \quad (2)$$

where $\varepsilon_1$ is the metal permittivity and $d$ its thickness, $\varepsilon_2$ the permittivity of the surrounding environment, $k_{z1,2}^2 = \varepsilon_{1,2}\left(\frac{\omega}{c}\right)^2 - k_x^2 = \left(\frac{\omega}{c}\right)^2\left(\varepsilon_{1,2} - \varepsilon_{eff}\right)$, and $c$ is the speed of light *in vacuo*. The solutions of Eqs. (1) and (2) can be found by minimizing the functions $\omega^+$ and $\omega^-$. The minimization procedure is accomplished by implementing a bi-dimensional Nelder-Mead algorithm [53]. This method allows us to find the complex wavevectors of the high and low frequency modes and, as a consequence, the effective permittivity values ($\varepsilon_{eff}$). High- and low-frequency modes may be then classified with respect to their propagation distance on the metal-dielectric interface. Long range surface plasmons (LRSPs) are characterized by a dispersion relation given by Eq. (1), relatively poor confinement [51], and have a relatively small imaginary part of $\varepsilon_{eff}$ as they are able to propagate longer distances on the metal surface. On the other hand, short range surface plasmons (SRSPs) (dispersion relation given by Eq.(2)) display a larger imaginary part of $\varepsilon_{eff}$, smaller mode size, and better overlap of the mode with the metal layer [54]. As the field invades the metal layer ohmic losses increase thus curtailing propagation distances.

We aim to understand the effects of SRSPs on transmission and absorption properties of metal gratings that contain ultra-thin metal bridges. We first calculate the effective permittivities of both SRSPs and LRSPs supported by thin, continuous metal layers, having thickness $d$, and surrounded by air ($\varepsilon_2$). In what follows we consider Ag as the metal of choice, with permittivity values ($\varepsilon_1$) taken from Ref. [55]. The main difference between SRSPs and LRSPs may be found in the values of their effective permittivities (both real and imaginary parts). As SRSPs are waves "bound" to the metal surface, both $\text{Re}(\varepsilon_{eff})$ and $\text{Im}(\varepsilon_{eff})$ assume values that are larger compared to those of the LRSPs for all wavelengths under investigation (see Fig. 1 (a) and (b)) and all thicknesses studied. LRSP waves extend more in the adjacent dielectric material (air in this case), taking on values much closer to the permittivity of this medium (see Fig. 1 (c) and (d)). The effective permittivity values shown in Fig. 1 may be used to evaluate the effective refractive index (both real and imaginary parts) of the same modes. This information, along with an evaluation of the effects of the truncation of the metal layer, clarifies the origin of transmission/absorption spectra modifications caused by the insertion of metal bridges inside the slits. A discussion of these resonant conditions is postponed to the next section.

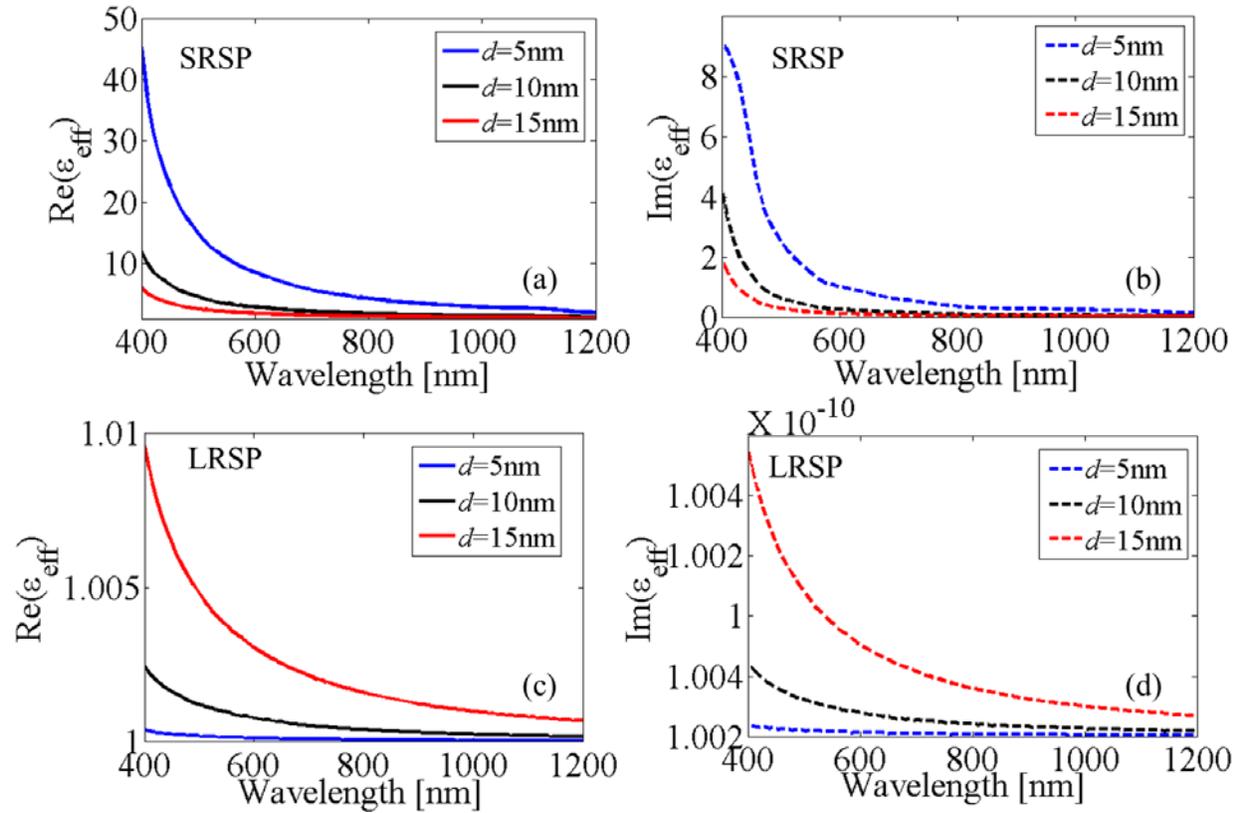

Figure 1: (a) Real and (b) imaginary part of the effective permittivity $\varepsilon_{eff}$ of SRSPs propagating on silver layers of thickness $d$. (c) Real and (d) imaginary part of $\varepsilon_{eff}$ of LRSP waves for silver layers of thickness $d$.

## 3. Thick metal gratings with ultra-thin bridges

We first analyze transmission/absorption spectra of a zero-order metal grating without thin bridges (Fig. 2 (a)) and then compare the same quantities by considering ultra-thin bridges that cap the bottom of the slit (Fig. 2 (b)). In what follows we assume a TM-polarized incident field, i.e., electric field lying in the plane of incidence. All calculations and results reported in this paper were obtained using two different computational methods that yield nearly identical results: (i) a full-wave approach based on the Finite Element Method (Comsol Multiphysics) [56] and (ii) a Rigorous Coupled Wave Analysis (RCWA) [57]. For this analysis we consider gratings with periodicity $p$ = 350 nm, while the wavelength range of interest is 400 nm – 1200 nm. We illuminate the structure with a plane wave at normal incidence ($\theta$ = 90º in Fig. 2). This will prevent the onset of localized resonances for the bridge which are normally suppressed at normal incidence for TM polarization [58]. The choice of a small pitch value and normal incidence condition simplifies the analysis even further by avoiding resonant mechanisms introduced by the grating, i.e., Wood's anomalies[1,2] and plasmonic band gap[10]. In fact, Wood's anomalies occur at $\lambda_{WA} = p\sqrt{\varepsilon_2}/l$, while the plasmonic band gaps are centered at $\lambda_{BG} = p\sqrt{\varepsilon_1\varepsilon_2/(\varepsilon_1+\varepsilon_2)}/l$, with $l$ =1,2,…. Both first order ($l$ = 1) Wood's anomaly and plasmonic band gap occur below 400nm.

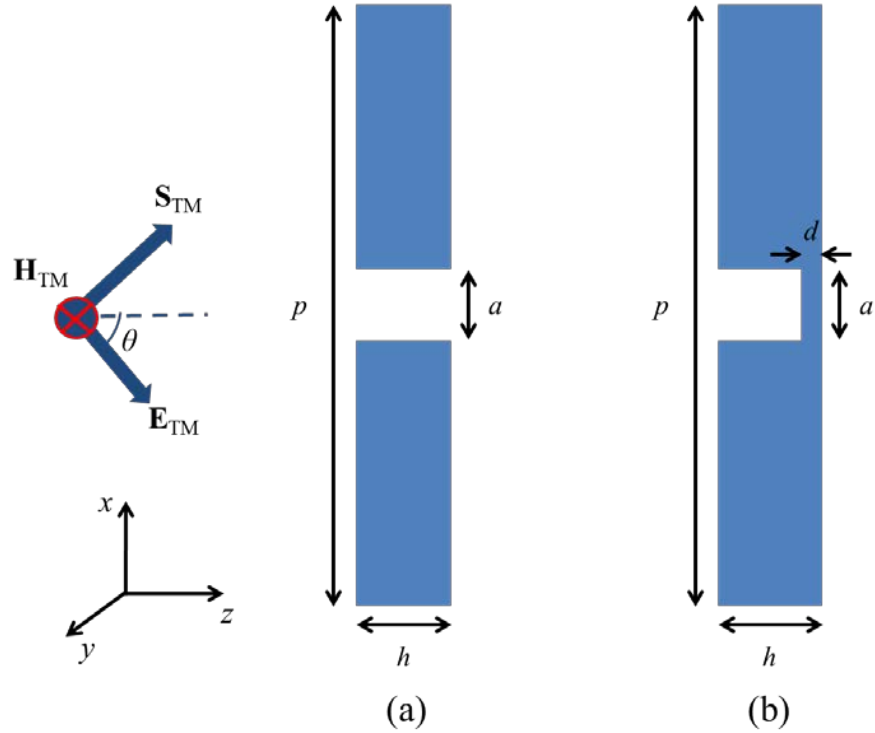

Figure 2: Sketch of the structures under investigation: (a) open grating with aperture $a$, thickness $h$, periodicity $p$ (b) grating with aperture $a$, thickness $h$ and periodicity $p$ capped by a thin metal bridge of thickness $d$.

Under these circumstances an open grating with subwavelength slits support only Fabry-Perot-like cavity resonances [10,15] that are governed by layer thickness $h$ and aperture size $a$. The first set of simulations (Fig.3) shows the dependence of the transmission/absorption peaks on aperture size, which is varied between 30 nm (air filling ratio ~8%) and 300 nm (air filling ratio ~86%). A 300 nm-thick grating supports two Fabry-Perot modes (bright regions in Fig. 3). However, the spectral position of these resonances depends on aperture size, as one may easily infer from Fig. 3 (a). No other resonant mechanisms characterize this grating.

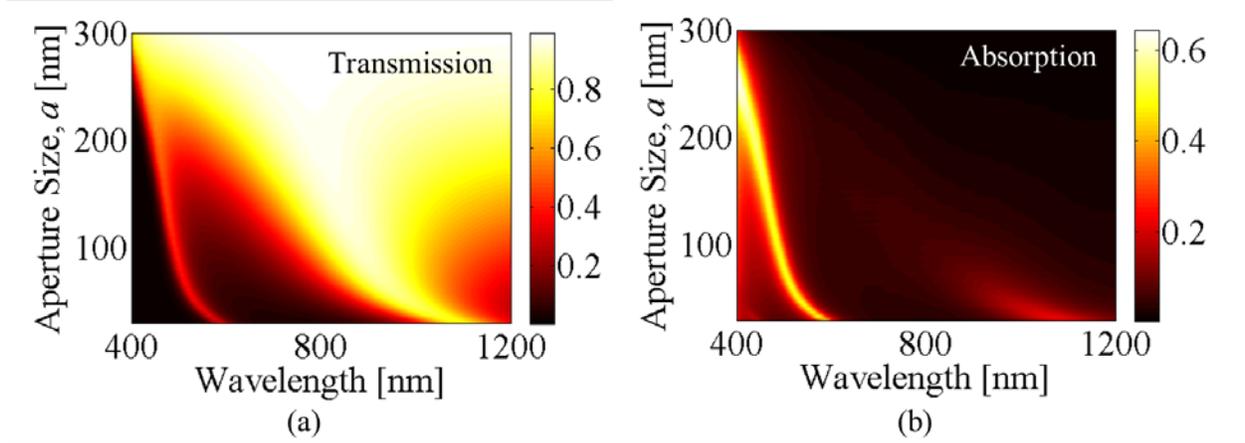

Figure 3: (a) Transmission and (b) absorption as a function of wavelength and aperture size at normal incidence for an open grating with thickness $h = 300$ nm and periodicity $p = 350$ nm (Fig.2 (a)).

We now consider a grating with the slits capped by a thin metal layer having thickness $d$. The grating is otherwise identical to the open grating (Fig. 2(b)). We analyze the absorption of the modified grating in four different scenarios where $d = 15$ nm, 10 nm, 5 nm, 2 nm (Fig.4). Aperture size and incident wavelength are varied as in the previous analysis (see Fig.3). We recognize that nonlocal and quantum size effects, and even fabrication techniques may enter the picture for bridge thicknesses below 5nm. For example, it has been shown that the imaginary part of the dielectric permittivity of silver nanoparticles increases when particle size decreases [59]. However, we presently assume that these effects are negligible and postpone their discussion to a future effort.

We first note that the closed slit causes a blue-shift of the Fabry-Perot resonances of the slits. One may associate this with a change of the reflection coefficient of one of the two mirrors of the nanocavity formed by the truncated MIM waveguide. In the open slit scenario the Fabry-Perot modes occur for cavity lengths $L_{FP} = \dfrac{q\lambda_0}{2n_{eff}}$, where $q$ is an integer, $\lambda_0$ is incident wavelength, and $n_{eff}$ the effective refractive index of the metal-air-metal waveguide [47,49]. In contrast, in a closed cavity scenario (thick or semi-infinite metal substrate closing the cavity and mimicking a perfect mirror) the standing wave oscillates in an open-ended, organ-pipe-like fashion [60], and the resonant lengths follow the equation $L_{FP\_CLOSED} = \dfrac{(2q-1)\lambda_0}{4n_{eff}}$ [39,40].

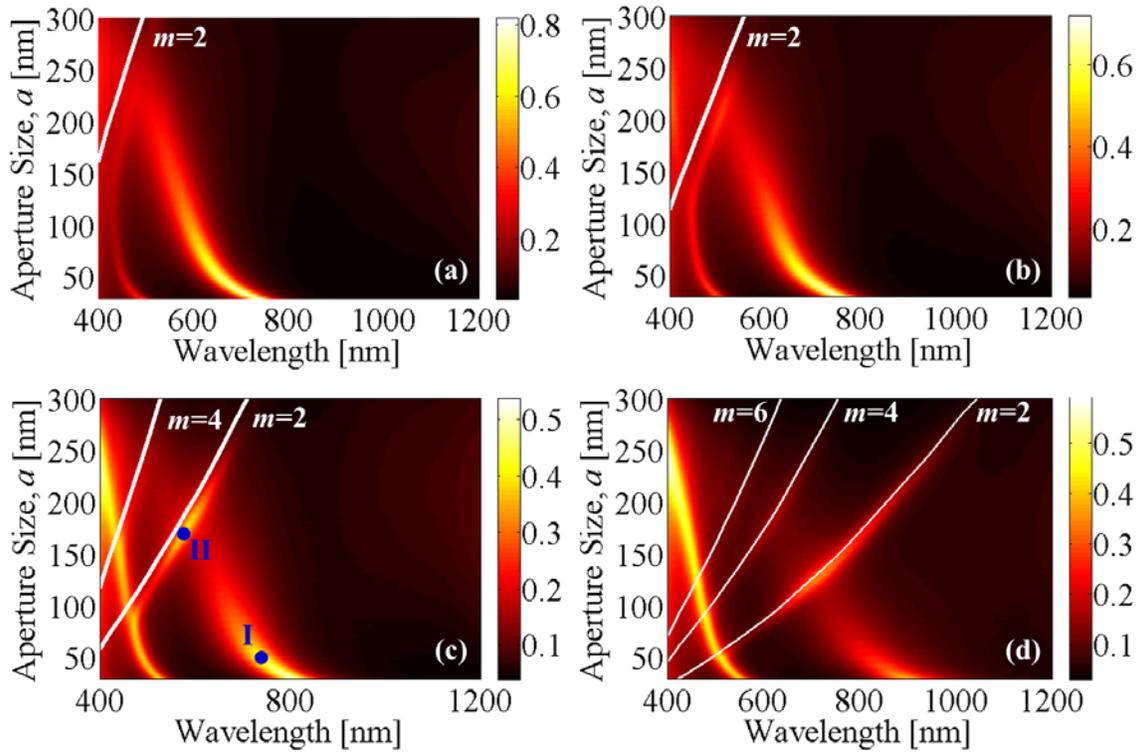

Figure 4: Absorption as a function of wavelength and aperture size at normal incidence for grating thickness $h = 300$ nm and periodicity $p = 350$ nm, closed with an ultra-thin metal bridge (see sketch in Fig.2 (b)) having thicknesses: (a) $d = 15$ nm, (b) $d = 10$ nm, (c) $d = 5$ nm and (d) $d = 2$ nm.

However, since here we are capping the cavity with a thin, semi-transparent metal bridge, one may generally observe a partial transition between closed- and open-cavity regimes as a function of bridge thickness. Closing the grating with a 15 nm-thick metal bridge (Fig. 4 (a)) causes the first order Fabry-Perot mode ($q = 1$) to dramatically blue-shift with respect to the open cavity case (Fig. 3 (b)), as the second order Fabry-Perot mode ($q = 2$) fades, as expected. In fact, only odd Fabry-Perot resonances ($q = 1, 3, 5, ...$) are supported in slits closed with a perfect conductor [40,39]. As the metal bridge becomes thinner ($d = 10$ nm in Fig. 4 (b), $d = 5$ nm in Fig. 4 (c) and $d = 2$ nm in Fig. 4 (d)), the absorption profile approaches the profile of the open cavity (Fig. 3 (b)). The intriguing and perhaps more relevant features that appear in the absorption maps in Fig. 4 are additional absorption peaks (marked by solid, white lines). To understand the origin of these extra peaks, we assume that if the metal bridge were removed from the grating it may be considered as a simple nano-antenna with resonant wavelengths related to its physical length. In contrast, the resonant conditions of an isolated nano-antenna may be altered by truncating the object [54,50], and retardation effects induced by such truncation increase proportionally to the difference between the effective index of the guided mode of the metal strip and the refractive index of the surrounding environment. Since the metal bridge here is partially truncated by metal walls with the same refractive index as the antenna, we may, as a first approximation, neglect such retardation effects and write the equation that governs the resonances of the thin metal bridge in a metal grating as:

$$a = \frac{m\lambda_0}{2\operatorname{Re}(n_{SRSP})} \quad . \quad (3)$$

where $a$ is the length of the antenna, i.e., the grating aperture, $m$ is an integer, $\lambda_0$ the incident wavelength, and $n_{SRSP}$ is the effective refractive index of the short range surface plasmon of the thin metal layer calculated from Eq. (2). Fig. 4 shows that the metal bridge placed in the slit of a metal grating resonates and produces regions of enhanced absorption when even modes are excited along the metal bridge ($m = 2, 4, 6,\ldots$). We also note that, as the metal bridge becomes thinner, the real part of the effective refractive index of the supported modes increases (see Fig.1), so that higher-order modes may be observed for the same aperture sizes (up to 3 modes when $d = 2$ nm – see Fig. 4 (d)). These resonant modes are strongly localized in the metal layer and effectively increase absorption up to ~60%. A near field plot (Fig.5) helps us to understand the substantial difference in field localization for different resonant conditions. In particular, Fig. 5 shows: I) a Fabry-Perot resonance, and II) the interference of a Fabry-Perot resonance with a localized bridge resonance ($m = 2$).

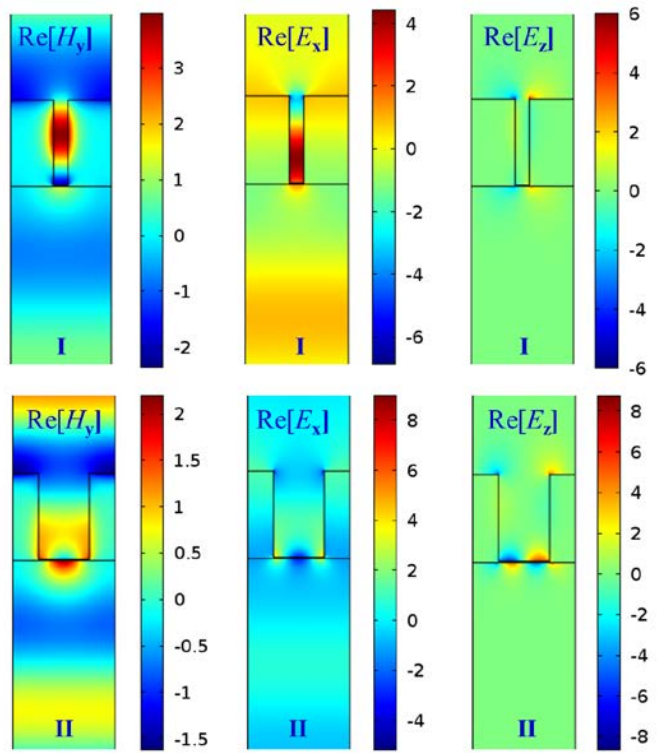

Figure 5: Plot of the real part of magnetic and electric field phasors for points I and II shown in Fig. 4(c). Point I refers to a thick grating with thickness $h = 300$ nm, periodicity $p = 350$ nm, $d = 5$ nm, $a = 50$ nm, and incident wavelength 735 nm. Point II refers to a thick grating with thickness $h = 300$ nm, periodicity $p = 350$ nm, $d = 5$ nm, $a = 175$ nm, and incident wavelength 580 nm.

It is well known that the bright modes of an isolated antenna illuminated at normal incidence display only odd parity ($m = 1, 3, 5,...$) [54,50,58] for symmetry reasons. However, when other resonators are present these resonant wavelengths may shift significantly [61,62] and lead to a change of the parity of the modes, eventually allowing even modes to be bright.

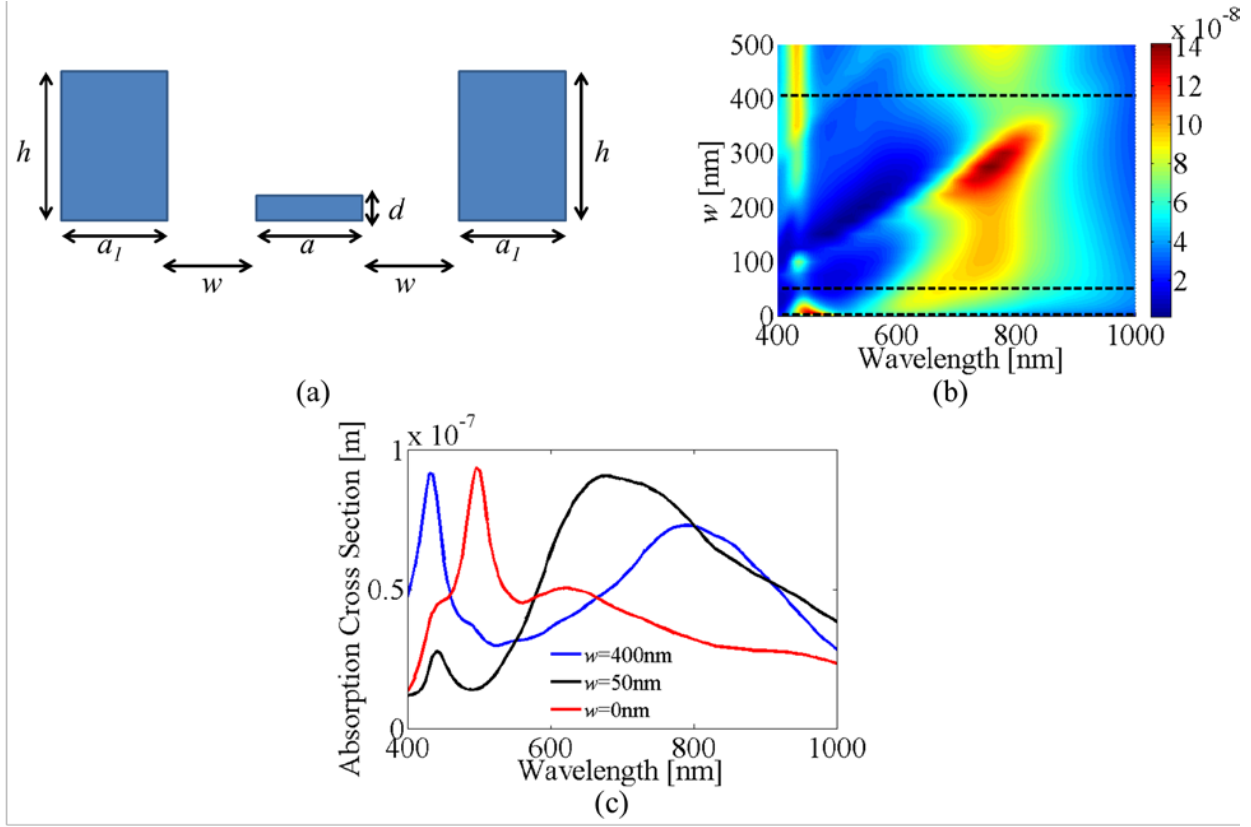

Figure 6: (a) Sketch of three resonators with sizes $a = 200$ nm, $a_1 = 150$ nm, $d = 10$ nm, $h = 300$ nm separated by a variable distance $w$; (b) Absorption cross section of the central resonator ($a = 200$ nm, $d = 10$ nm) as a function of the distance $w$; (c) Sections of the map in (b) for $w = 400$ nm (blue line), $w = 50$ nm (black line) and $w = 0$ nm (red line).

The transition of bright modes from odd to even symmetry, which takes effect completely only when the resonators are in contact ($w = 0$ in Fig.6 (b)), may be monitored by plotting the absorption cross section of the metal bridge ($d = 10$ nm, $a = 200$ nm) when a larger resonator ($h = 300$ nm, $a_1 = 150$ nm) is placed on either side (see sketch in Fig. 6 (a)) at variable distance $w$. Fig. 6(b) shows how two resonances (corresponding to the odd modes $m = 1$ and $m = 3$) are discernible when the larger resonators are far ($w > 400$ nm) from the central one. The same resonances broaden and come closer, eventually merging into a single resonant state when the resonators touch ($w = 0$). For clarity three sections of the map in Fig. 6(b) (dashed, black lines) are shown in Fig. 6(c). The sections help to visualize the broadening, shifting and merging of the resonant states of the central resonator sketched in Fig. 6(a), as the resonators combine into a single nanocavity. We note that the field distribution of the single resonance observed when $w = 0$ (red line in Fig. 6(c)) corresponds to a mode with even parity ($m = 2$) localized in the bridge with thickness $d$.

## 4. Thin metal grating with ultra-thin bridges

The distinction between the thick and thin metal regimes is pivotal. Thin metal gratings are attractive because they can suppress transmission [63-65], which in turn may mean enhanced reflection and/or absorption. Although the excitation of the bridge resonance is not contingent on the presence of Fabry-Perot-like resonances inside the slit (with or without cap), the effects of these modes on the transmission/absorption spectra may vary dramatically depending on the thickness of the grating. As the grating becomes thinner, Fabry-Perot resonances of the slits are cutoff and the grating begins to support localized resonances. While the former event is not critical, the latter complicates the situation as bridge and grating resonances begin to interact. However, the transition from the *thick grating regime* (discussed in the previous section) to the *thin grating regime* (discussed in this section) is gradual and may be monitored by varying the grating thickness for a fixed aperture size. The steepness of this transition and the thickness for which the localized resonances of the bridge and the grating begin to interfere may depend on aperture size and bridge thickness. As an example here we plot the absorption map for a grating with $a = 175$ nm, $p = 350$ nm, and $d = 4$ nm (Fig. 7(a)). The map reveals that bridge resonances (white lines labeled $m = 2$ and $m = 4$ described by Eq. (3)) are always present, regardless of the presence of Fabry-Perot modes inside the slit. Fabry-Perot resonances are the thickness-dependent bright regions observed only for $h > 70$ nm, undergoing a red shift as $h$ increases (Fig.

7(a)). On the other hand, we observe that the enhanced absorption regime induced by the excitation of bridge resonances is altered when the grating localized resonances become stronger ($h < 40$ nm) (Fig. 7(b)).

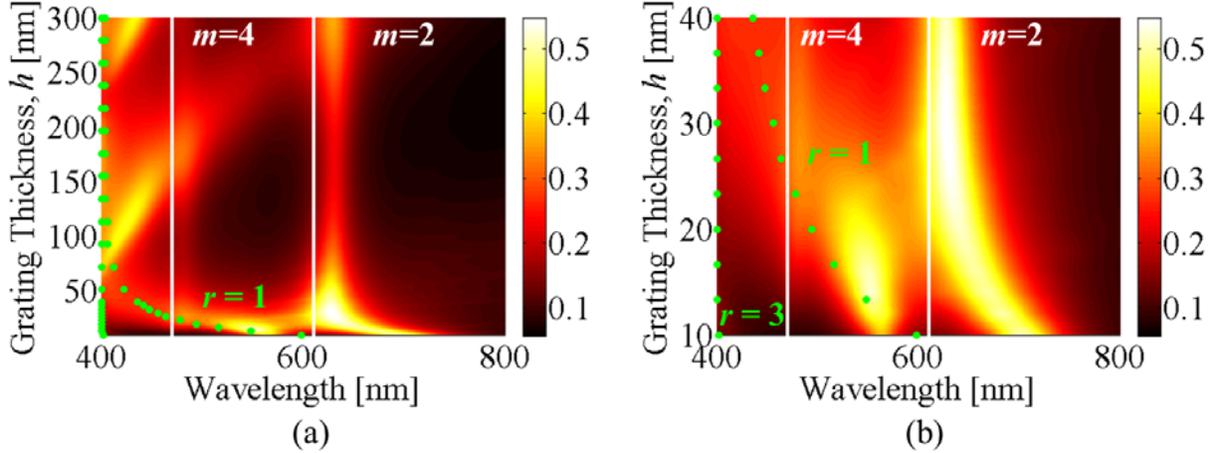

Figure 7: (a) Absorption as a function of wavelength and grating thickness ($h$) at normal incidence for a grating with periodicity $p = 350$ nm, aperture $a = 175$ nm, closed with an ultra-thin metal bridge (see sketch in Fig.2 (b)) with thickness $d = 4$ nm; (b) zoom of (a) for $h < 40$ nm.

A thin, open grating can support localized resonances, since it is equivalent to a periodic arrangement of nano-antennas in air. Without taking into account the reflection phase shift introduced by the abrupt truncation of the silver strip [54], one may approximate the dispersion curve of the localized resonances of the grating by means of the equation:

$$p - a = \frac{r\lambda_0}{2\,\mathrm{Re}(n_{SRSP})}, \quad (4)$$

where $p$-$a$ is the length of the antenna, $r$ is an integer, $\lambda_0$ the incident wavelength, and $n_{SRSP}$ is the effective refractive index of the short range surface plasmon of the thin metal layer calculated using Eq. (2). By plotting the solution of Eq. (4) for every thickness $h$ and mode order $r$ (green circles labeled $r = 1$ and $r = 3$ in Fig.7) we observe that a modulation of absorption occurs at the intersection of the localized bridge and grating resonances.

The interplay of these resonances may be monitored more clearly as a function of the aperture size and wavelength as was done for the thick grating. For example, a zero-order, open, silver grating (Fig. 2(a)) with $h = 10$ nm and $p = 350$ nm can suppress transmission (dark regions in Fig. 8(a)) and enhance absorption (bright regions in Fig. 8(b)). The resonators are surrounded by air, and only odd modes are permitted (solid green curves, Fig. 8(b), solution of Eq. (4) for $r = 1$ and $r = 3$). We now close the grating with an ultra-thin metal bridge (as sketched in Fig. 2(b)) having thickness $d$. Differently from the case shown in Sec.3, where the open grating was not able to support any localized resonance, we now observe a dramatic interaction between the localized modes of the grating (dispersion curves described by Eq. (4) – solid green curves in Fig. 9) and the localized modes of the bridge (dispersion curves described by Eq. (3) – solid white curves in Fig. 9). Several anti-crossing regions occur whenever resonant modes with different parity interact (odd modes supported by the thicker resonator and even modes supported by the bridge).

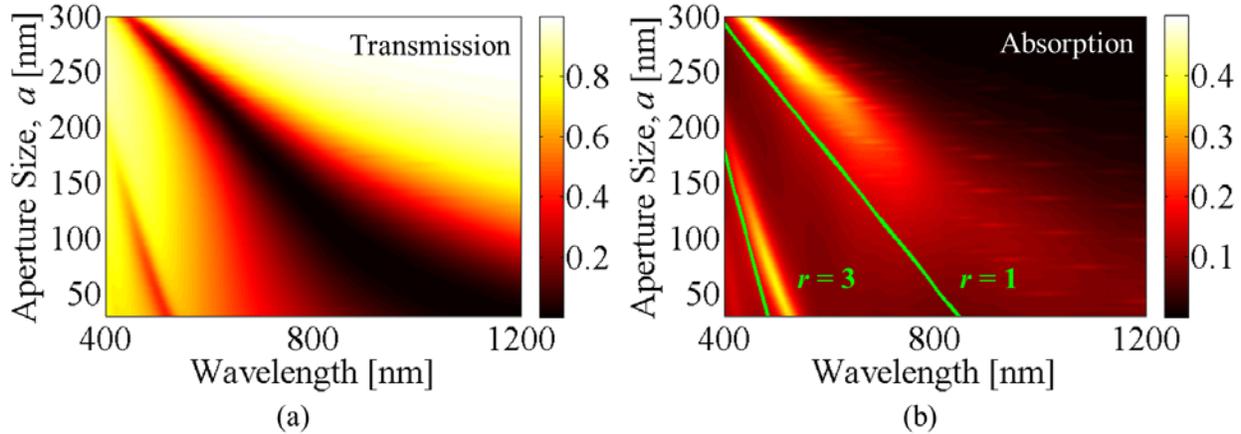

Figure 8: (a) Transmission and (b) absorption as a function of wavelength and aperture size at normal incidence for an open grating with thickness $h = 10$ nm and periodicity $p = 350$ nm (see sketch of the structure in Fig.2 (a)).

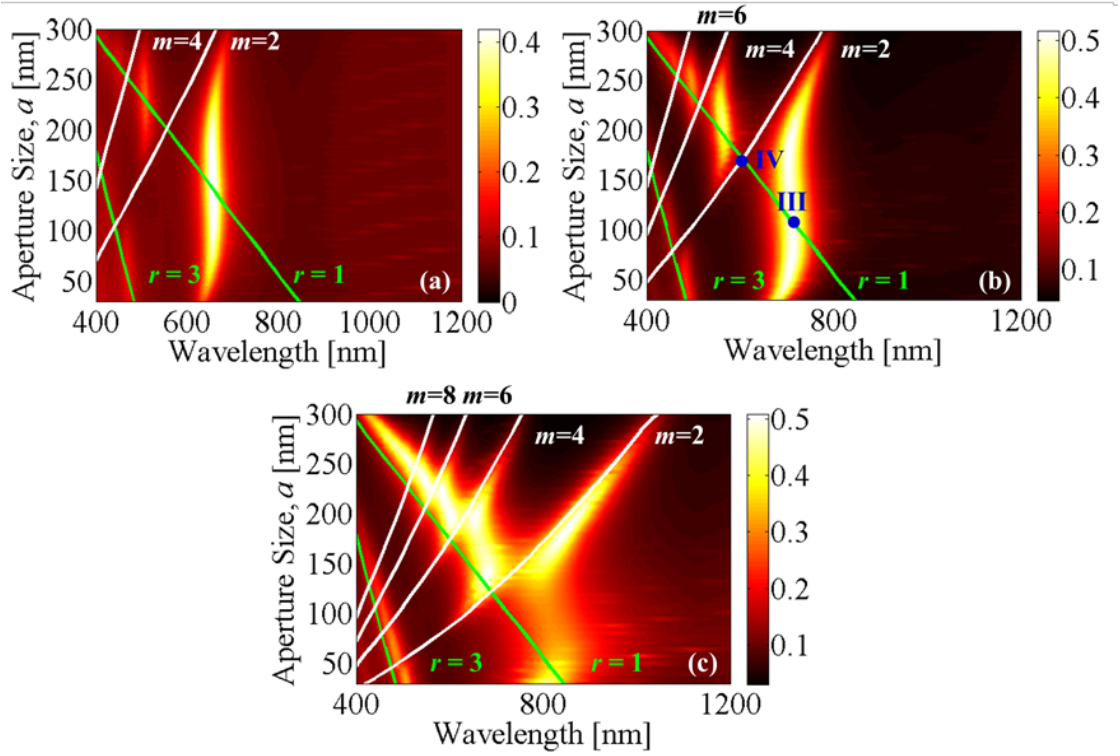

Figure 9: Absorption as a function of wavelength and aperture size at normal incidence for a grating with thickness $h$ = 10 nm and periodicity $p$ = 350 nm, closed with an ultra-thin metal bridge (see sketch in Fig.2 (b)) with thickness (a) $d$ = 6 nm, (b) $d$ = 4 nm, and (c) $d$ = 2 nm.

The result is a decrease in electric field amplitude, as also shown in other systems with multiple resonances [66,67], and consequent suppression of absorption, which is proportional to $\mathrm{Im}(\varepsilon)|E|^2$. At the same time, absorption increases with respect to the open grating case (Fig. 8) outside the anti-crossing regions, achieving values near 50%. One may also conclude that under specific circumstances distortion and splitting of the resonances lead to narrow absorption bands that are virtually independent of aperture size (see for example Fig. 9(a) and (b)). This peculiar behavior suggests the possibility of realizing filters that are insensitive to fabrication defects.

Just as we did for the thick-grating scenario, we now plot (see Fig. 10) the near field components for two different conditions (marked as points III and IV in Fig. 9(b)). Point III helps to visualize field distribution for a grating localized resonance, while point IV shows how the fields localize in the anti-crossing region, revealing how the structure becomes almost completely transparent thanks to the interference of a grating localized resonance ($r = 1$) and a localized bridge resonance ($m = 2$) (see Fig. 10).

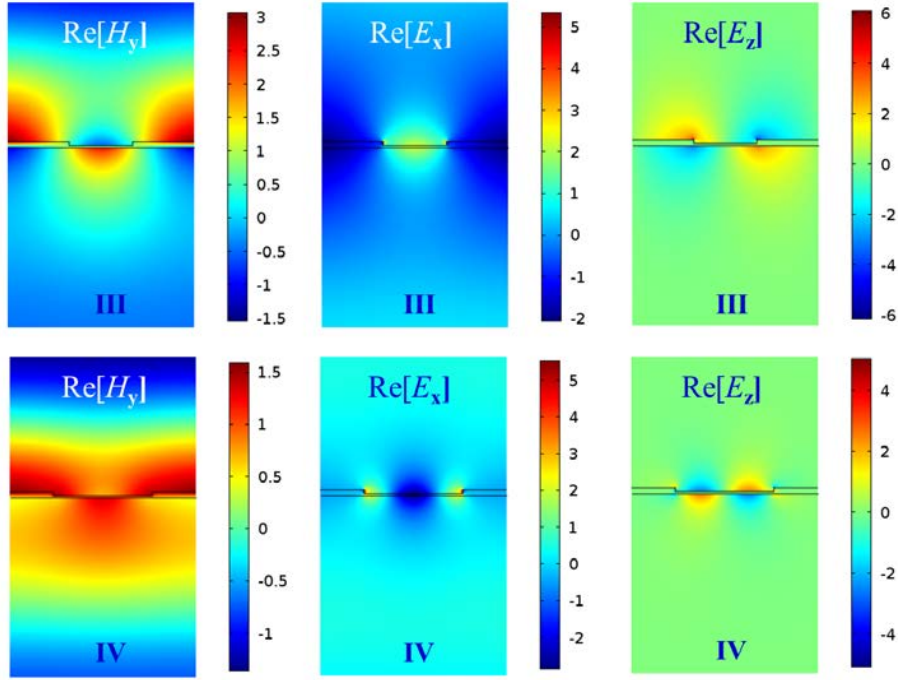

Figure 10: Plot of the real part of magnetic and electric field phasors for points III and IV shown in Fig. 9(b). Point III refers to a thin grating with thickness $h = 10$ nm, periodicity $p = 350$ nm, $d = 4$ nm, $a = 110$ nm, and incident wavelength 708 nm. Point IV refers to a thin grating with thickness $h = 10$ nm, periodicity $p = 350$ nm, $d = 4$ nm, $a = 175$ nm, and incident wavelength 608 nm.

## 5. Role of the position of the bridge inside the slits

It is known that plasmonic elements like resonant metallic nanoparticles [68,69] introduced inside the slits of a lamellar grating can modify transmission and absorption spectra of the grating. However, the consequences of placing resonant, plasmonic nanostructures inside the slits *in contact* with the metal grating have not been discussed. The presence of an extremely thin metal layer that caps the grating may be either intentional or accidental. For example, when focused ion-beam fabrication processes are not optimized, or when adhesion layers are not removed efficiently during lithographic processes, one should expect changes to the spectral response of the structure. On the other hand, if one is interested in creating such ultra-thin bridges intentionally, it may be useful to know whether or not the position of the bridge is critical to achieve certain spectral properties. For this reason in what follows we simulate two situations, one with a metal bridge placed in the *center* of both thick and thin gratings (as shown in Fig. 11), the other with an equivalent scenario where the bridge is placed at the *edge* of the slit (as shown in Fig. 2(b)).

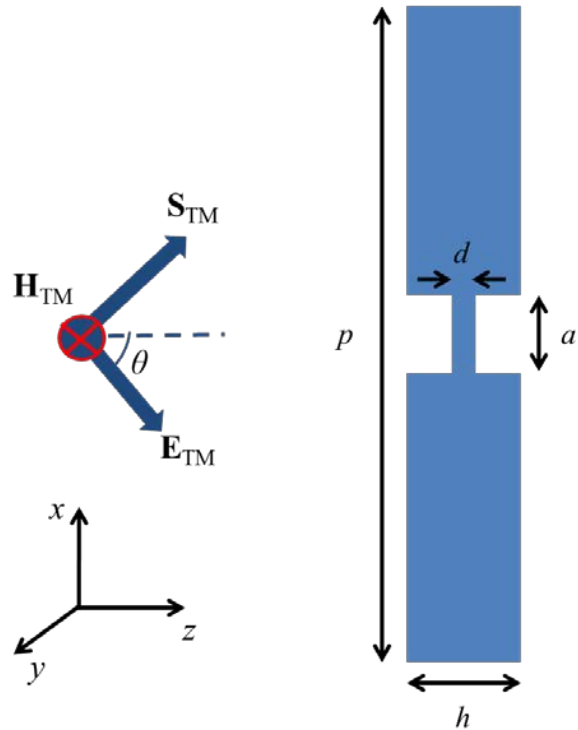

Figure 11: Sketch of the structure with aperture *a*, thickness *h* and periodicity *p* with a thin metal bridge having thickness *d* placed in the middle of the slit.

In previous sections we have seen that thick and thin metal gratings behave differently with respect to the excitation of localized resonances on metal bridges. For this reason in what follows we will discuss the role of the position of the metal bridge in these two regimes separately. In a thick grating scenario, transmission and absorption spectra are dominated mostly by Fabry-Perot resonances. An interruption inside the slit or a cap at the end of the slit may in fact alter significantly the Fabry-Perot resonances' spectral positions [68] or the parity of these modes [39,40]. In order to compare the effects of the position of the bridge in a scenario where Fabry-Perot resonances are supported, i.e., thick gratings, we calculate the absorption profile as a function of aperture size and incident wavelength for a thick grating ($h$ = 300 nm, $p$ = 350 nm, $a$ = 200nm) with an ultra-thin bridge ($d$ = 10 nm) placed at the *edge* (Fig. 12 (a)) or at the *center* of the slit (Fig. 12(b)). By comparing Fig. 12(a) and Fig. 12(b) one may infer that the position of the bridge inside the slit: (i) does not modify the dispersion of the localized resonance and (ii) alters the spectral positions of Fabry-Perot resonances. For example, considering an aperture size $a$ = 30 nm, the first order Fabry-Perot resonance shifts from ~ 800 nm when the bridge is on the *edge* to ~ 1250 nm when the bridge is in the center (see Fig. 12(a) and (b)).

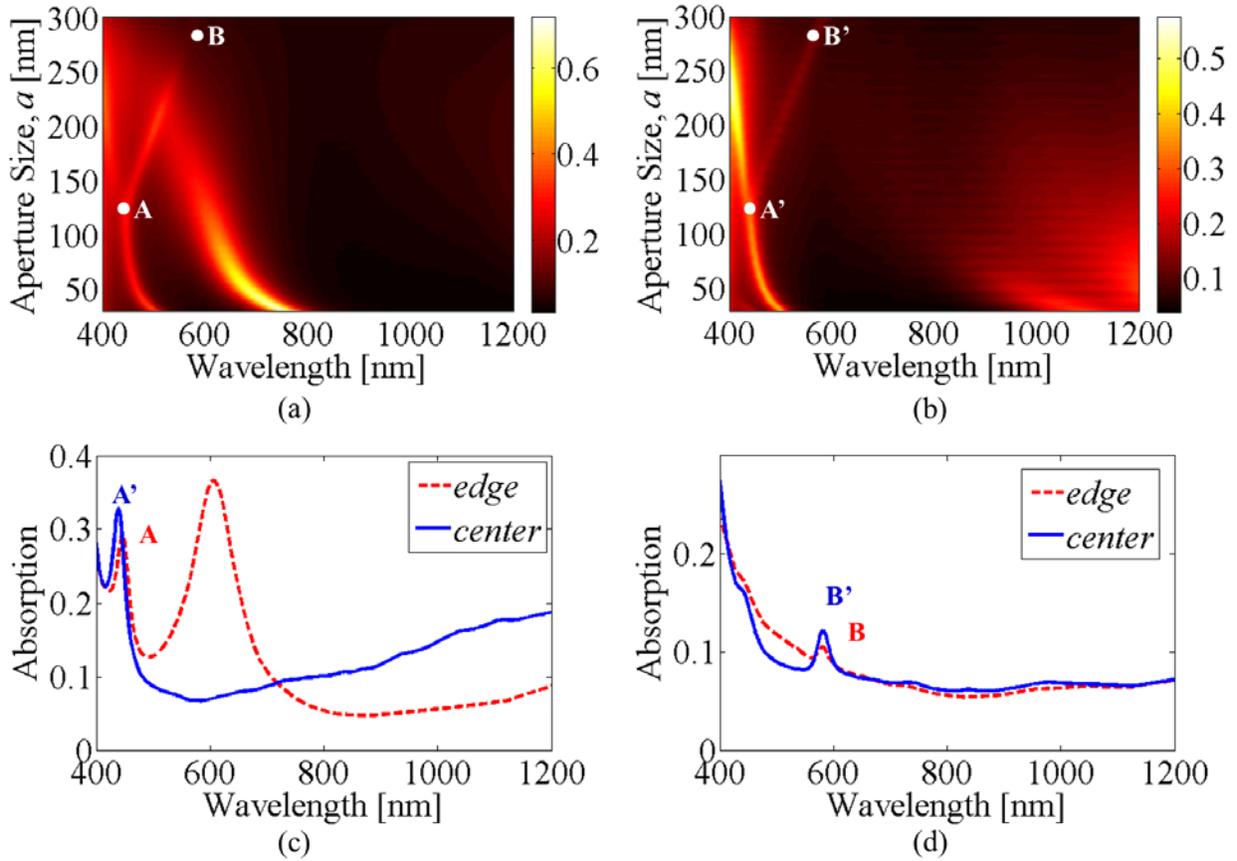

Figure 12: Absorption maps for a thick grating with variable aperture size, $h = 300$ nm, $p = 350$ nm and a ultra-thin metal bridge $d = 10$nm thick placed (a) at the *edge* of the slit and (b) in the *center* of the slit. Absorption spectra of two sections of (a) and (b) for (c) $a = 120$ nm and (d) $a = 298$ nm.

However, the absorption induced by the localized resonance of the metal bridge remains almost unaffected by the position of the bridge either in the presence (point A and A' in Fig. 12 (a) and (b)) or absence (point B and B' in Fig. 12 (a) and (b)) of a Fabry-Perot mode inside the slit. . Additionally, we compared two sections of the maps in Fig. 12(a) and (b). Fig. 12(c) shows the absorption profile for an aperture $a = 120$ nm: the peaks marked A and A', which exhibit similar values of absorption, are in the crossing points of a localized resonance branch and the second-order Fabry-Perot branch. Similarly, we obtain comparable absorption values (Fig. 12 (d)) for the peaks marked B and B' ($a = 298$ nm), which belong only to a localized resonance branch. However, we observe that the simultaneous excitation of a Fabry-Perot mode for the slit and a localized resonance of the bridge increases absorption.

If we now consider a thin grating scenario ($h = 10$ nm, $a = 200$ nm, $p = 350$ nm) the position of a 4 nm-thick metal bridge in the center of the slit, has no effect on the absorption profile (compare solid, blue line and red, dashed line in Fig. 13 (b)).

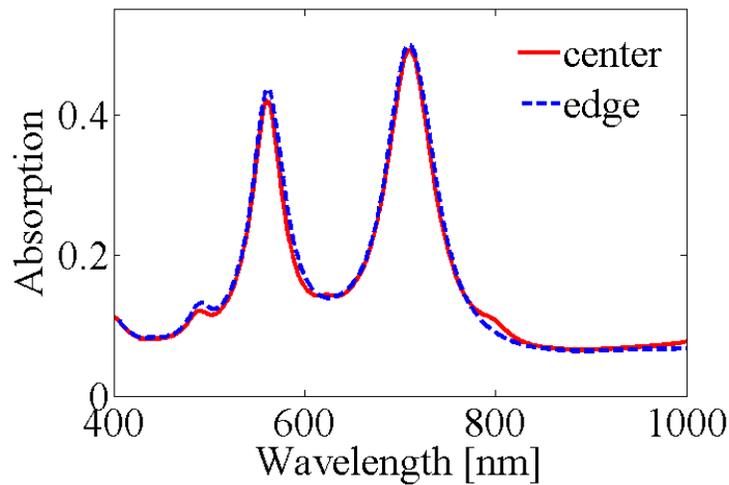

Figure 13: Absorption as a function of wavelength at normal incidence for a grating with thickness $h = 10$ nm, aperture $a = 200$ nm and periodicity $p = 350$ nm; a metal bridge $d = 4$ nm thick is placed in the center of the slit (red, solid line) or at the edge of the slit (blue, dashed line).

## 6. Conclusions

We have shown that ultra-thin metal bridges placed in the slits of lamellar gratings can support localized resonances for short range surface plasmon polaritons. The formation of these resonant modes alters dramatically the absorption properties of the grating. Two different operating regimes may be distinguished, depending on the ability of the lamellar structure to support localized resonances. If a bridge resonance is excited on a thick grating, the result is enhanced absorption, regardless of the position of the bridge inside the slit. In contrast, lamellar structures already able to support localized resonances without the bridge, i.e., thin gratings, exhibit anti-crossing and significant bending of absorption bands that become nearly insensitive to the geometry of the slit and the position of the bridge. The excitation of resonant modes inside ultra-thin metal regions with significant increase of the electric field in sub-wavelength regions is promising for nonlinear applications, where large order susceptibility values may be exploited to enhance third order nonlinear processes.

## Acknowledgement

This research was performed while the authors M. A. Vincenti and D. de Ceglia held a National Research Council Research Associateship award at the U.S. Army Aviation and Missile Research Development and Engineering Center. M. Grande thanks the U.S. Army International Technology Center Atlantic for financial support (Contract no. W911NF-12-1-0292). The authors also thank F. Dioguardi for helpful discussions.